\documentclass[aps,pra,reprint,showpacs]{revtex4-1}
\usepackage{amsbsy}
\usepackage{amssymb}
\usepackage{amsmath}
\usepackage{graphicx}
\usepackage[T1]{fontenc}
\usepackage[latin9]{inputenc}
\usepackage{esint}
\begin{document}

%\preprint{}


\title{Slow-light dissipative solitons in an atomic medium assisted by an incoherent pumping field}


\author{Nuno A. Silva}
%\thanks{}

\affiliation{Departamento de F\'{i}sica e Astronomia da Faculdade de Ci\^{e}ncias
da Universidade do Porto, Rua do Campo Alegre 687, 4169-007 Porto,
Portugal.}
\affiliation{INESC TEC, Centre of Applied Photonics, Rua do Campo Alegre
687, 4169-007 Porto, Portugal.}

\author{A. L. Almeida}
%\thanks{}

\affiliation{Departamento de F\'{i}sica e Astronomia da Faculdade de Ci\^{e}ncias
da Universidade do Porto, Rua do Campo Alegre 687, 4169-007 Porto,
Portugal.}
\affiliation{INESC TEC, Centre of Applied Photonics, Rua do Campo Alegre
687, 4169-007 Porto, Portugal.}

\author{Ariel Guerreiro}
\affiliation{Departamento de F\'{i}sica e Astronomia da Faculdade de Ci\^{e}ncias
da Universidade do Porto, Rua do Campo Alegre 687, 4169-007 Porto,
Portugal.}
\affiliation{INESC TEC, Centre of Applied Photonics, Rua do Campo Alegre
687, 4169-007 Porto, Portugal.}


\date{\today}

\begin{abstract}
This work models the propagation of an optical pulse in a 4-level
atomic system in the electromagnetic induced transparency regime.
By demonstrating that linear and nonlinear optical properties can
be externally controlled and tailored by a continuous-wave control
laser beam and an assisting incoherent pump field, it is shown how
these media can provide an excellent framework to experimentally explore
pulse dynamics in the presence of non-conservative terms, either gain
or loss. Furthermore, we explore the existence of stable dissipative
soliton solutions, testing the analytical results with computational
simulations of both the effective (1+1)-dimensional model and the
full Maxwell-Bloch system of equations.
\end{abstract}

%\keywords{}

\maketitle

\section{Introduction}

Laser-induced coherent effects between atomic states laid a cornerstone
in optics in the latest decades. Offering a plethora of effects, from
electromagnetically induced transparency\cite{EIT_OBSERVATION,EIT_review,EIT_REVIEW_PROGRESS}
to light-storage\cite{EIT_SLOWLIGHT,EIT_STORAGE,EIT_STORAGE_2}, the
potential of quantum optical systems to constitute an important building
block in future technologies has been extensively explored\cite{QUANTUM_NONLINEAROPTICS,TECHN_QUANTUM,TECH_QUANTUMOPTICS_PHOTONIC,TECH_PHOTONIC_SIMULATOR}.
Moreover, quantum enhanced nonlinearities characteristic of the near
resonance regime\cite{EIT_Nonlinear,EIT_SLOWLIGHT,N_NONLINEAR,N_LIQUIDLIGHT}
make these systems ideal for exploring nonlinear effects, which span
from the study of the dynamics of temporal\cite{TEMPORALSOLITON_1,TEMPORALSOLITONS,TEMPORALSOLITONS2}
and spatial solitons\cite{N_NONLINEAR,LAMBDA_LINLAT_KONOTOP,SPATIALSOLITONS_HONG,SPATIAL_SOLITONS_4level}
to the realization of optical analogue experiments\cite{NAS_PERSISTENT,FERREIRA_SUPER,SUPER_AOS_PRECONDENSATION}.
Still, most of these studies often overlook the non-equilibrium aspects
of the physical system. These include the interplay between dissipative
effects (that come from relaxation processes), the gain mechanisms
(arising from a pump optical beam for example), and even of the temporal
response and transient regime of atomic populations. Understanding
the role played by each of these aspects in different situations is
a key challenge for future theoretical and experimental work.

On what concerns nonlinear optics, optical solitons are a hallmark,
finding technological applications on communications and all optical
computing\cite{SOLITONS_OPTICS_BOOK,SOLITONS_OPTICS_COMM_2,SOLITONS_OPTICS_REVIEW_COMM,SOLTIONS_OPTICS_COMPUTING,SOLITONS_COMPUTING}.
Resulting from a balance between dispersion and nonlinearity, their
stable shape is usually lost if the optical media features dissipation
and gain. Remarkably, even in such non-equilibrium conditions it is
still possible to observe a class of soliton solutions. These solutions,
called dissipative solitons or auto-solitons result from a double
dynamical equilibrium between dispersion and nonlinearity and also
between dissipation and gain mechanisms\cite{AUTOSOLITONS_BOOK,DISSIPATIVESOLITONS_BOOK,DISSIPATIVESOLITONS_OPTICS_PAPER,LAMBDA_DissipativeSolitons_Coupled,LAMBDA_DISSIPATIVESOLITONS}.
While known to exist for a wide diversity of optical systems, in the
particular case of coherent media the existence and dynamics of dissipative
solitons is still under-explored\cite{LAMBDA_DissipativeSolitons_Coupled,LAMBDA_DISSIPATIVESOLITONS}.
In a recent study\cite{LAMBDA_DISSIPATIVESOLITONS}, it was predicted
the existence of temporal dissipative solitons in a hollow core photonic
crystal fiber filled with a $3$-level atomic system. However, in
that case, the dissipative properties of the system can only be varied
within a small range due to their dependence on the structure of the
fiber, which can be detrimental for an experimental observation of
the theoretical predictions. Moreover, the results were only confirmed
through numerical simulation of an effective model and not through
complete simulations of the Maxwell-Bloch equations\cite{MB_derivation_nonlinear,MB_numer_doppler,MB_sims_cross,DENSITY_MATRIX_BOOK},
which go well beyond the validity of the effective model and approximate
the simulations to real experimental conditions.

In this paper, we propose an experimentally-friendlier\textbf{ }alternative,
which allows to tune the linear and nonlinear gain and loss exploiting
the coherent optical properties of a 4-level atomic system assisted
by an incoherent pumping field\cite{INCOHERENT_4,INCOHERENTPUMP_grating}.
In addition, we extend the previous state of research by confirming
the existence of a family of temporal dissipative solitons through
numerical simulations of both the effective and the complete Maxwell-Bloch
model, and discuss the small discrepancies between the two models. 

This work is organized as it follows: first, in section II, we model
the optical response of a 4-level $N$-type atomic media confined
in a waveguide to a pulsed probe field, while driven simultaneously
by a coherent control electromagnetic field and an incoherent pump
field. We apply the standard multiscale perturbation technique\cite{TEMPORALSOLITONS}
to derive an effective one-dimensional Cubic Ginzburg-Landau equation
(CGLE), which describes the dynamics of the envelope of a pulsed probe
beam. In section III, we introduce the solutions for temporal dissipative
solitons previously proposed in the literature\cite{DS_sols_plasma,LAMBDA_DISSIPATIVESOLITONS}
and in section IV we explore the role of the incoherent pumping for
controlling the linear and nonlinear optical properties and the gain
and loss processes. After the correct choice of the parameters, in
section V we test a family of dissipative solitons by performing both
numerical simulations of the (1+1)-dimensional effective model but
also of the Maxwell-Bloch system. Finally, section VI presents our
conclusions.

\section{Physical model}

This work considers a waveguide filled with a 4-level atomic system
with\emph{ N}-type configuration interacting with two optical beams
driving the transitions between the levels in the near-resonant regime.
As illustrated schematically in Fig. 1., a weak pulsed probe field
$\boldsymbol{E}_{p}=\frac{\boldsymbol{e}_{p}}{2}\left[E_{p}\left(r,t\right)e^{i\left(\beta_{p}z-\omega_{p}t\right)}+\mbox{c.c.}\right]$
drives the transition $\left|1\right\rangle \rightarrow\left|3\right\rangle $
with envelope function $E{}_{p}\left(r,t\right)$, propagation constant
$\beta_{p}$ and polarization vector $\boldsymbol{e}_{p}$ and a center
frequency $\omega_{p}$ corresponding to a detuning $\Delta_{p}\equiv\omega_{p}-\left(\omega_{3}-\omega_{1}\right)$.
The transition between levels $\left|2\right\rangle \rightarrow\left|3\right\rangle $
is driven by a continuous-wave strong control field $\boldsymbol{E}_{c}=\frac{\boldsymbol{e}_{c}}{2}\left[E_{c}\left(r\right)e^{i\left(\beta_{c}z-\omega_{c}t\right)}+\mbox{c.c.}\right]$,
with envelope function $E{}_{c}\left(r\right)$, wave vector $\beta_{c}$
and polarization vector $\boldsymbol{e}_{c}$ and a center frequency
$\omega_{c}$ corresponding to a detuning $\Delta_{c}=\omega_{c}-\left(\omega_{3}-\omega_{2}\right)$,
which for simplicity we choose to satisfy $\Delta_{c}=\Delta_{p}$,
that stands\textbf{ }for the Raman resonance condition and usually
minimizes the absorption. Additionally, a two-way incoherent field
with pumping rate $P$ drives the transition $\left|1\right\rangle \rightarrow\left|4\right\rangle $.
Considering that the polarizations are orthogonal, $\boldsymbol{e}_{p}\cdot\boldsymbol{e}_{c}=0$,
the evolution of the probe beam can be described by the wave equation
\begin{equation}
\partial_{z}^{2}\boldsymbol{E}_{p}+\nabla_{\perp}^{2}\boldsymbol{E}_{p}-\frac{1}{c^{2}}\partial_{t}^{2}\boldsymbol{E}_{p}=\frac{1}{\varepsilon_{0}c^{2}}\partial_{t}^{2}\boldsymbol{P}.\label{eq:waveq}
\end{equation}
Here, the polarization term acts a source and accounts for the optical
response of the atomic medium, defined by taking the dipolar approximation
as 
\begin{equation}
\boldsymbol{P}=\bar{\eta}e^{i\left(\beta_{p}z-\omega_{p}t\right)}\boldsymbol{\mu}_{31}\rho_{31}+\mathrm{c.c.},\label{eq:polarization}
\end{equation}
where $\bar{\eta}$ is the averaged atomic density, $\boldsymbol{\mu}_{31}$
is the dipole moment of the transition $\left|3\right\rangle \rightarrow\left|1\right\rangle $
and $\rho_{ij}$ are the matrix elements of the density matrix operator
$\boldsymbol{\rho}$. On the other hand, the dynamics of the density
operator is described by the master equation

\begin{equation}
\dot{\boldsymbol{\rho}}=-\frac{i}{\hbar}\left[\hat{H},\boldsymbol{\rho}\right]-\frac{\hat{\Gamma}\left(\boldsymbol{\rho}\right)}{2},\label{eq:Master_equation}
\end{equation}
where $\hat{\Gamma}$ is the Lindblad superoperator describing the
relaxation and dephasing processes, $H$ is the Hamiltonian of the
system given by
\begin{eqnarray}
\hat{H} & = & \sum_{i=1}^{4}\hbar\omega_{i}\left|i\right\rangle \left\langle i\right|-\frac{\hbar}{2}\left(\Omega_{p}e^{-i\omega_{p}t}\left|3\right\rangle \left\langle 1\right|\right.\nonumber \\
 &  & \left.\Omega_{c}e^{-i\omega_{c}t}\left|3\right\rangle \left\langle 2\right|+\mbox{H.c.}\right),\label{eq:2hamiltonian}
\end{eqnarray}
where the $\Omega_{i}$ are the Rabi frequencies for the transitions
defined as $\Omega_{p}=\boldsymbol{e}_{p}\cdot\boldsymbol{\mu}_{31}E_{p}/\hbar$
and $\Omega_{c}=\boldsymbol{e}_{c}\cdot\boldsymbol{\mu}_{32}E_{c}/\hbar$.
Taking the rotating wave approximation, we eliminate the rapid oscillatory
terms and derive the explicit form of equation \ref{eq:Master_equation},
as detailed in the supplementary material\cite{Suplemental}. 

To solve the Maxwell-Bloch system(MB) that is formed by equations
\ref{eq:waveq} and \ref{eq:Master_equation}, we use the standard
multiscale approach, which produces approximated results valid for
different time scales and avoids the occurrence of potentially diverging
secular terms\cite{MULTIPLESCALES_TEMPORAL,LAMBDA_DISSIPATIVESOLITONS}.
This technique consists in defining the multiscale variables $t_{l}=\delta^{l}t$
and $z_{l}=\delta^{l}z$ and introducing the series expansion of the
envelope function and density operator 

\begin{eqnarray}
E_{p} & = & \sum_{l=1}\delta^{l}{\cal E}_{p}^{\left(l\right)},\quad\boldsymbol{\rho}=\sum_{l=0}\delta^{l}\bar{\boldsymbol{\rho}}^{\left(l\right)}\label{eq:multiscale-2}
\end{eqnarray}
in equations \ref{eq:waveq} and \ref{eq:Master_equation}. Then,
an hierarchy of equations is obtained by separating the equations
into their dependence on different orders of the parameter $\delta$.
After some cumbersome calculations outlined in the supplementary materials\cite{Suplemental},
we obtain a equation for the Rabi frequency $\Omega_{p}\left(z,\tau\right)$
as

\begin{multline}
i\partial_{z}\Omega_{p}+\frac{\beta_{p}^{''r}}{2}\partial_{\tau}^{2}\Omega_{p}+g^{r}\left|\Omega_{p}\right|^{2}\Omega_{p}=\\
\beta_{p}^{'i}\partial_{\tau}\Omega_{p}-i\beta_{p}^{i}\Omega_{p}-ig^{i}\left|\Omega_{p}\right|^{2}\Omega_{p}-i\frac{\beta_{p}^{''i}}{2}\partial_{\tau}^{2}\Omega_{p},\label{eq:NLSEcomplex derivation-1}
\end{multline}
where $\tau=t-\beta_{p}^{'r}z$, $\beta_{p}=\beta_{p}^{r}+i\beta_{p}^{i}$
is the propagation constant, and $g=g^{r}+ig^{i}$ is related with
the nonlinear susceptibility of the atomic media. The two remaining
parameters $\beta_{p}^{'}=\beta_{p}^{'r}+i\beta_{p}^{'i}$ and $\beta_{p}^{''}=\beta_{p}^{''r}+i\beta_{p}^{''i}$
are second-order and third-order expansion terms, respectively, and
are related to the group velocity and to the group velocity dispersion
as $v_{g}\equiv1/\beta_{p}^{'^{r}}$ and $\mbox{GVD}\equiv1/\beta_{p}^{''^{r}}$.
Furthermore, we have separated each of the complex-valued parameters
into their real and imaginary parts, which detach the conservative
terms(associated with the real part) from the non-conservative ones(that
can be either loss or gain and are associated with the pure imaginary
parts).

\begin{figure}
\begin{centering}
\includegraphics[width=8.6cm]{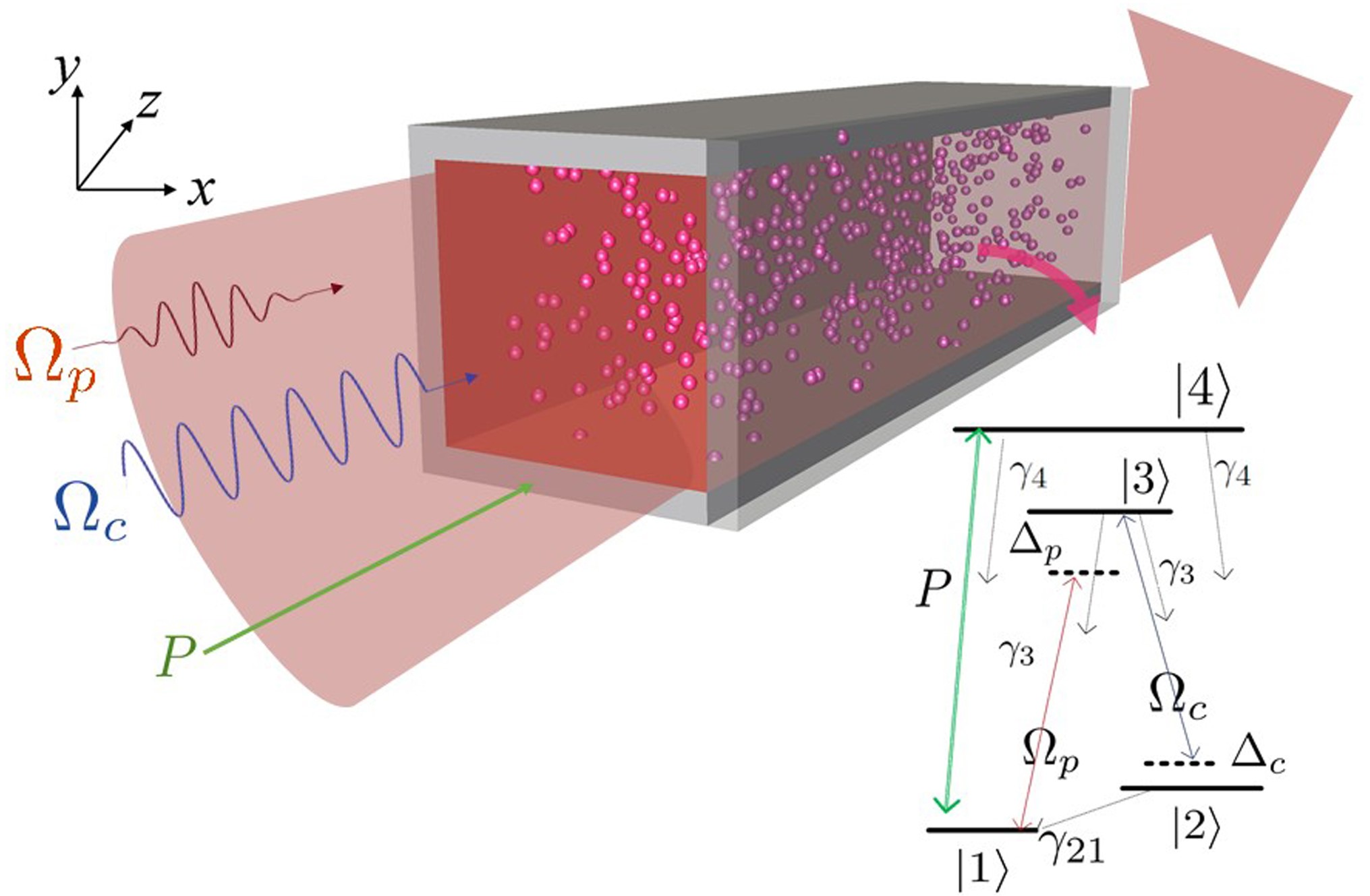}
\par\end{centering}

\caption{Illustration of the physical problem under study, where a rectangular
waveguide filled with the atomic gas is interacting with two coherent
optical beams. A diagram of the energy levels for the atomic gas is
also depicted. Each of transitions is driven by an electromagnetic
field, a pulsed probe $\Omega_{p}$, a continuous wave strong control
$\Omega_{c}$. Also represented are the detunings, the decoherence
rates $\gamma_{i}$ and the incoherent pumping field $P$.}
\end{figure}

All these quantities have an intricate dependency on the physical
properties of the atomic system, which are discussed in more detail
in the supplementary material\cite{Suplemental}. Still, we emphasize
that they depend on the characteristics of the physical system (specifically
on the decay and dephasing rates), as well as, on a set of experimentally
tunable parameters, namely the incoherent pumping rate $P$ and the
detunings of the coherent fields $\Delta_{p}$ and $\Delta_{c}$.
Thus, in principle, the optical response of the atomic system can
be tailored externally in real experiments, which allows to explore
the existence of dissipative solitons in regions of the space parameters
where the double dynamical equilibrium between dispersion and nonlinearity
and dissipation and between gain processes is achieved.

\section{Dissipative soliton solutions}

The multiscale approach introduced in the last section and outlined
in the supplemental material\cite{Suplemental} allowed to transform
the Maxwell-Bloch equation system into a simplified (1+1)-dimensional
model, which correspond to the well-known Complex Ginzburg-Landau
Equation(CGLE). By rescaling $\tau\rightarrow\tau/\sqrt{\left|\beta_{p}^{''r}\right|}$
and $\Omega_{p}\rightarrow\sqrt{\left|g^{r}\right|}\Omega_{p}$, we
can get the adimensional form of the former equation written as

\begin{multline}
i\frac{\partial\Omega_{p}}{\partial z}+\frac{s_{d}}{2}\frac{\partial^{2}\Omega_{p}}{\partial\tau^{2}}+s_{g}\left|\Omega_{p}\right|^{2}\Omega_{p}\\
=i\alpha\Omega_{p}+i\zeta\frac{\partial^{2}\Omega_{p}}{\partial\tau^{2}}+i\epsilon\left|\Omega_{p}\right|^{2}\Omega_{p}+\xi\frac{\partial\Omega_{p}}{\partial\tau},\label{eq:CGLE}
\end{multline}
with $s_{d}\equiv\beta_{p}^{''r}/\left|\beta_{p}^{''r}\right|$ and
$s_{g}\equiv g^{r}/g^{r}$ where $\sigma\left(q\right)\equiv q/\left|q\right|$,
is the signal function, $\alpha=-\beta_{p}^{i}$, $\zeta=-\beta_{p}^{''i}/\left(2\left|\beta_{p}^{''r}\right|\right)$,
$\epsilon=-g^{i}/\left|g^{r}\right|$ and $\xi=\beta_{p}^{'i}/\sqrt{\left|\beta_{p}^{''r}\right|}$.
The CGLE can be found in a myriad of physical systems and qualitatively,
the left hand-side terms of the equation describe the dynamics of
an envelope function in a nonlinear medium, while the extra terms
in the right hand side account for the dissipative or gain phenomena.

Neglecting the terms on the right hand side of equation \ref{eq:CGLE},
it reduces to the well studied Nonlinear Schrodinger equation, which
admits stable localized wave solutions called solitons when an equilibrium
between dispersion and nonlinear effects is achieved. Depending on
the signals of the dispersion and of the nonlinearity, these can be
either of the bright type (if $\sigma\left(\beta_{p}^{''^{r}}\right)\sigma\left(g^{r}\right)=1$)
or dark type (if $\sigma\left(\beta_{p}^{''^{r}}\right)\sigma\left(g^{r}\right)=-1$).
Usually in optics, the bright solitons are more relevant, and for
the standard Nonlinear Schrodinger equations the family of solutions
can be simply expressed as
\begin{equation}
\Omega_{p}\left(z,\tau\right)=a\mbox{sech}\left[a\left(z-v\tau\right)\right]\exp\left(ivz-i\left(a^{2}-v^{2}\right)\tau/2\right)\label{eq:soliton-1-1-1}
\end{equation}
where $a$ and $v$ are both free parameters associated with the amplitude
and the velocity, respectively. However, in the presence of the additional
terms on the right-hand side, these solutions become unstable. Remarkably,
it has been proven that equation \ref{eq:CGLE} can still admit a
special type of soliton-like solutions called dissipative solitons
or autosolitons, if conditions of equilibrium between gain and loss
are met, in addition to the balance between dispersion and nonlinearity\cite{DS_sols_1s,DS_sols1996,DS_sols_Analitical,DS_sols_CQ,DS_sols_plasma}.

In a recent study, Fac\~{a}o and co-workers have investigated the existence
of temporal optical dissipative solitons in a gas-filled fiber with
a 3-level atomic media\cite{LAMBDA_DISSIPATIVESOLITONS}. The family
of solutions proposed is an extension of those previous predicted
in the literature for cases where $\beta_{p}^{i}=0$\cite{DS_sols_plasma},
and are given by
\begin{equation}
\Omega_{p}\left(z,\tau\right)=A\mbox{sech}\left[B\left(\tau-\frac{z}{v}\right)\right]^{1+id}\exp\left(iCz-iD\tau\right)\label{eq:soliton-1-1}
\end{equation}
where 
\begin{eqnarray}
A & = & \sqrt{\frac{3B^{2}d\left(1+4\zeta^{2}\right)}{2\left(2s_{g}\zeta-s_{d}\epsilon\right)}}\nonumber \\
B & = & \sqrt{\frac{\alpha+\frac{\xi^{2}}{4\zeta}}{\zeta d^{2}-\zeta+s_{d}d}}\nonumber \\
C & = & B^{2}\left[2\nu d-\frac{s_{d}\left(d^{2}-1\right)}{2}\right]-s_{d}D^{2}\label{eq:ds_parameters}\\
D & = & -\frac{\xi}{2\nu}\nonumber \\
v & = & s_{d}\frac{2\nu}{\xi},\nonumber 
\end{eqnarray}
and with the chirp parameter $d$ equals to 
\[
d=\frac{-3\left(1+2\epsilon\zeta\right)+\sqrt{8\left(2\zeta s_{g}-s_{d}\epsilon\right)^{2}-9\left(1+2\epsilon\zeta\right)}}{2\left(2\zeta s_{g}-s_{d}\epsilon\right)}.
\]
It is important to notice that contrary to what happens with the non-dissipative
case, the parameters are now fixed by the optical properties of the
system, and this family of solutions is only stable\cite{LAMBDA_DISSIPATIVESOLITONS}
provided that the following conditions are met simultaneously 
\begin{eqnarray}
\zeta & > & 0\label{eq:condition1}\\
\alpha+\frac{\xi^{2}}{4\zeta} & > & 0\label{eq:condition2}\\
\epsilon & > & \frac{\zeta\left(3\sqrt{1+4\zeta^{2}}-1\right)}{4+18\zeta^{2}}.\label{eq:condition3}
\end{eqnarray}
However, it happens that while the soliton solution is stable \emph{per
se} if these conditions are satisfied, it turns out that they coincide
exactly to the instability conditions for the background, which occurs
in the complementary of condition (\ref{eq:condition2}), i.e., for
$\alpha+\frac{\xi^{2}}{4\zeta}<0$, as it was previous noted in other
studies on analytical solutions of the CGLE\cite{DS_sols1996,DS_sols_CQ}.
This means that the stability of the solution is limited by the growth
rate of the background, which can be proven that to be significant
at distances $z_{ins}\sim\left(\alpha+\frac{\xi^{2}}{4\zeta}\right)^{-1}$,
thus constituting a boundary for the propagation of these dissipative
soliton solutions. 

In the previous work\cite{LAMBDA_DISSIPATIVESOLITONS}, Fac\~{a}o et.
al propose that the linear loss usually present in a 3-level atomic
media can be transformed into a gain by exploiting the transverse
confinement of light by the structure of the hollow crystal fiber,
which means that it is dependent of structural design of the system
and therefore not controllable. In the next section, we demonstrate
how the use of an additional incoherent pumping in the 4-level atomic
system introduced in this paper can result in an experimentally friendlier
setup to explore the phenomenology of temporal dissipative solitons
in optical systems, as it allows an easier control of the optical
properties of the system on a wider parameter range, which turns out
to be also helpful to control the stability of the solution.

\section{Tuning the optical properties}

As discussed in section II, the optical properties of the system can
be controlled externally by an appropriate choice of experimentally
tunable parameters, namely the incoherent pumping rate $P$ and the
detunings of the coherent fields $\Delta_{p}$. Furthermore, the 4-level
system proposed here can be realized with a multitude of atomic species
and just to give a practical example of real experimental values,
we consider for now the hyperfine structure of the D line of $^{85}\mbox{Rb}$
atoms\cite{steck2001rubidium} filling a rectangular waveguide of
dimensions $L_{x}=L_{y}=10\mbox{\ensuremath{\mu}m}$ at a fixed atomic
concentration $\eta=10^{12}\mbox{cm}^{-3}$. Assigning the hyperfine
structure levels $5S_{1/2}\left(F=1\right)$, $5S_{1/2}\left(F=2\right)$,$5P_{1/2}\left(F=2\right)$
and $5P_{3/2}\left(F=1\right)$ to $\left|1\right\rangle $, $\left|2\right\rangle $,$\left|3\right\rangle $
and $\left|4\right\rangle $, respectively, the dipole matrix elements
are given by $\mu_{13}\simeq6.74\times10^{-30}\mbox{Cm}$ and $\mu_{23}\simeq2.24\times10^{-30}\mbox{Cm}$.
Moreover, at typical room temperature experiments, the decoherence
rates are approximately the same for every excited level, meaning
that $\gamma_{3}\approx\gamma_{4}\approx\gamma_{2}\approx\gamma=3,6\times10^{7}\mbox{s}^{-1}$.
Finally, we will also consider the case for which the control beam
has a constant spatial intensity with $\left|\Omega_{c}\right|=\gamma$.

As we are interested in the observation of dissipative solitons of
the family proposed in section III, we shall now look to the regions
of the space parameters $\left(P,\Delta_{p}\right)$ that fulfill
the conditions (\ref{eq:condition1}-\ref{eq:condition3}) (herein
referred as conditions A) as well as $\sigma\left(\beta_{p}^{''^{r}}\right)>0\land\sigma\left(g^{r}\right)>0$
(herein conditions B), while maximizing $z_{ins}\sim\left(\alpha+\frac{\xi^{2}}{4\zeta}\right)^{-1}$
in order to limit the effects of the background instability. Figure
2a) represents the regions of the parameters space where the conditions
are met while Figure 2b) shows the dependence of the magnitude of
$z_{ins}$ in the region where both conditions are satisfied. A graphical
analysis of Figure 2b), together with a numerical search on the parameters
space suggests that $z_{ins}$ is maximized near the boundary of the
intersection of the two conditions, which happens to coincide also
with the line defined by the condition $\beta_{p}^{i}=0$. Remarkably,
this case corresponds to the absence of linear absorption/gain and
therefore to a case of perfect electromagnetic induced transparency.
This characteristic is uncommon and distinct from what is observed
with the most common 3-level $\Lambda$ atomic system, which usually
features a linear absorption due to the dephasing rate between levels
$\left|2\right\rangle $ and $\left|1\right\rangle $. As previously
suggested in the literature\cite{INCOHERENTPUMP_grating}, the suppression
of the normal linear optical absorption in the 4-level system can
be attributed to a gain mechanism introduced by the incoherent pumpimg
to the additional level. While not the main result of this work, this
interesting and unusual response is important for the observation
of dissipative solitons, but can also be relevant for the study of
other problems, including the case of spatial dissipative solitons\cite{DS_SPATIAL_INHOMO,INCOHERENTPUMP_grating}.

Considering a control beam intensity with $\left|\Omega_{c}\right|=4\gamma$,
the choice of detuning and the incoherent pumping rate as $\Delta_{p}=4\gamma$
and $P=0.174\gamma$ respectively, together with the expressions given
in the supplementary material\cite{Suplemental}, allow us to obtain
$\delta=0$, $\zeta=1.49$, $\xi=1.55$ and $\epsilon=-5.42$ with
$z_{ins}\approx2.4$. Therefore, this choice of parameters corresponds
to a system which features no linear gain/absorption and only includes
effects of nonlinear absorption and spectral filtering\cite{LAMBDA_DISSIPATIVESOLITONS}.
Moreover, for this set of values we obtain a group velocity for the
pulse of $0.3\%$ of the vacuum speed of light. As discussed in the
previous section, if the conditions A and B are met, it is possible
to observe a class of dissipative soliton solutions that we will test
in the next section through numerical simulations.

\begin{figure}
\begin{centering}
\includegraphics[width=8.6cm]{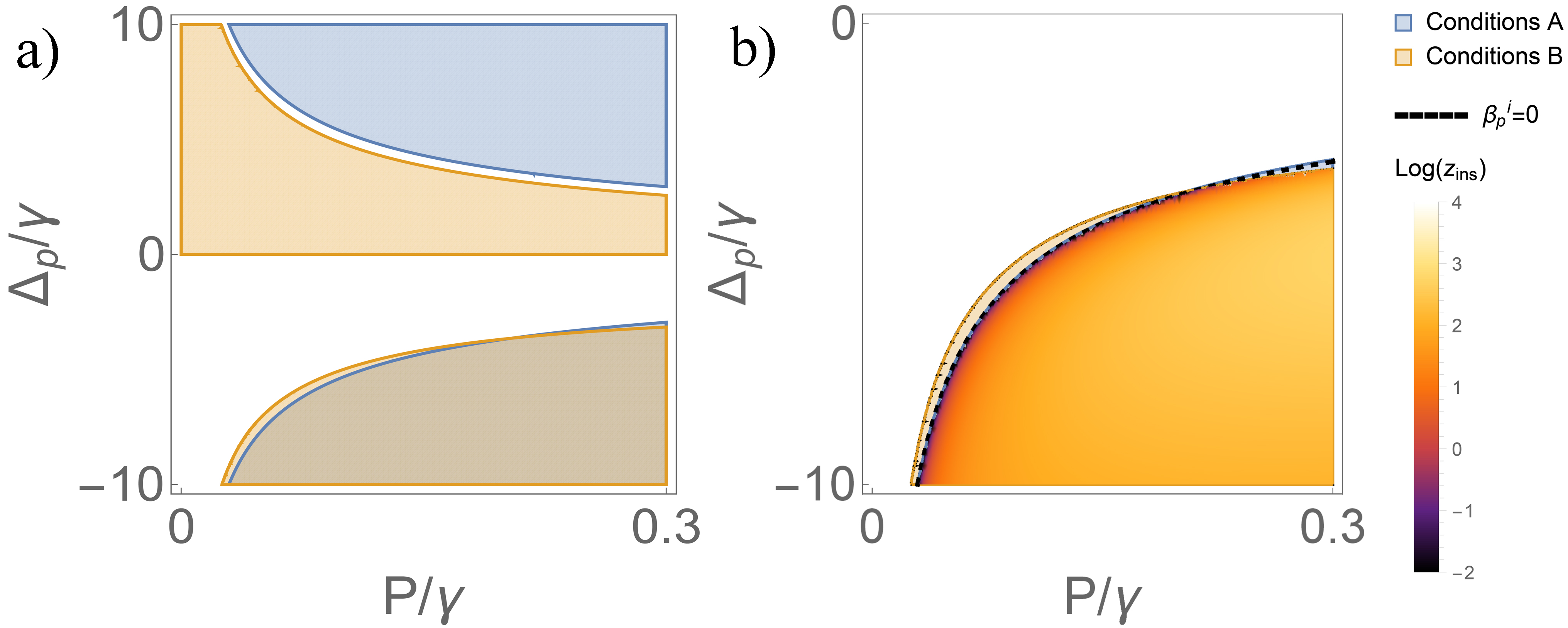}
\par\end{centering}

\caption{Illustration of the physical problem under study, where a rectangular
waveguide filled with the atomic gas is interacting with two coherent
optical beams. The diagram of the energy levels for the atomic gas
is also depicted. Each of transitions is driven by an electromagnetic
field, a pulsed probe $\Omega_{p}$, a continuous wave strong control
$\Omega_{c}$. Also represented are the detunings, the decoherence
rates $\gamma_{i}$ and the incoherent pumping field $P$.}
\end{figure}

\section{Numerical results and Discussion}

To test the existence and stability of dissipative solitons in the
atomic system proposed section IV, we have performed numerical simulations
of both the full MB and the effective CGLE models. In both cases,
we used high performance numerical tools based on general purpose
GPU programming frameworks, an approach already employed in previous
works\cite{NAS_PERSISTENT,SILVA_PRA_BEC,MB_comp_JPC,FERREIRA_SUPER}.
The MB system was solved using a leap-frog method and a finite-differences
scheme\cite{MB_comp_JPC}, while the CGLE solver uses a standard beam
propagation technique based on the split-step Fourier method(SSFM)\cite{SILVA_PRA_BEC}. 

\begin{figure}
\begin{centering}
\includegraphics[width=8.6cm]{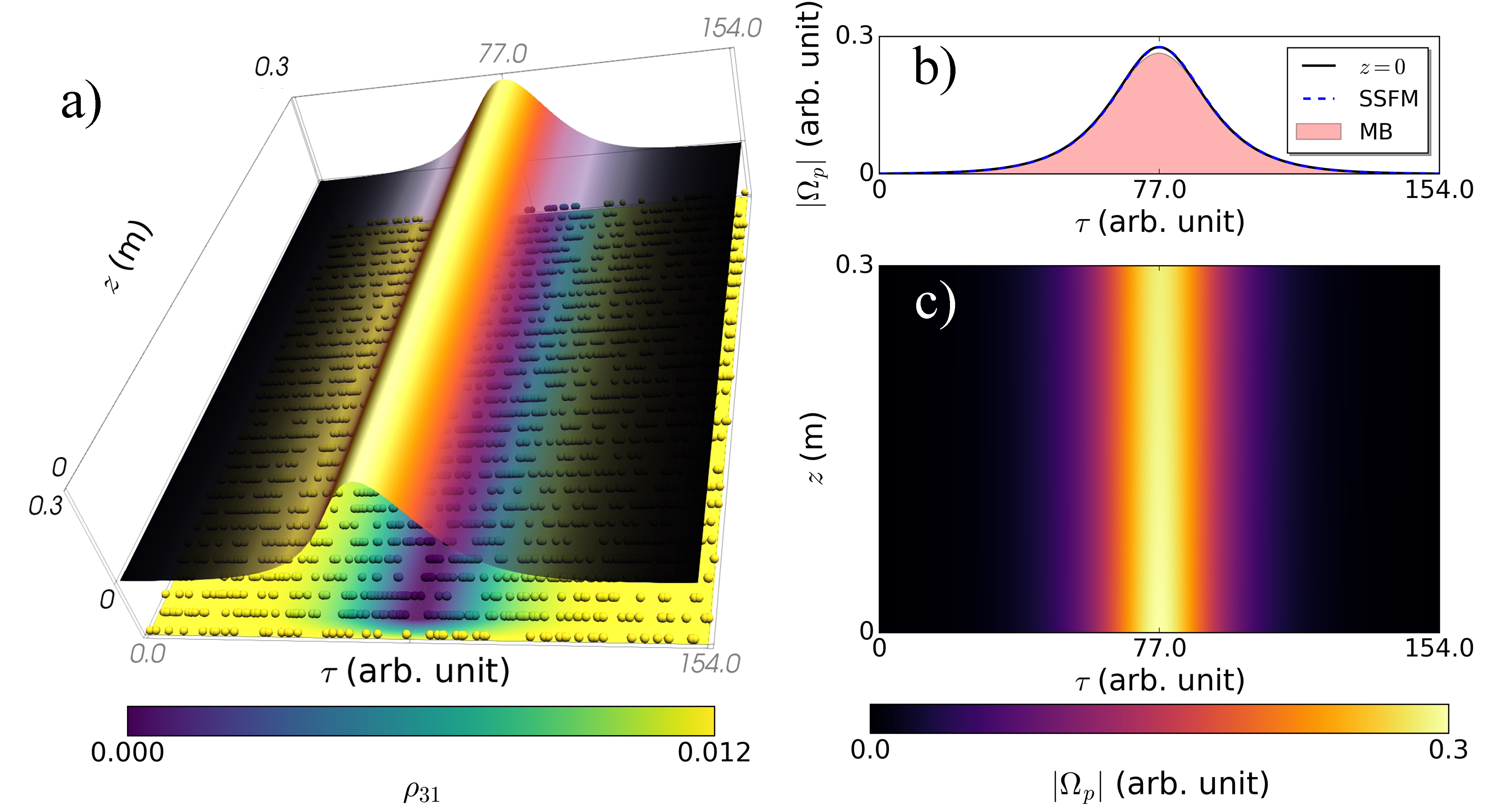}
\par\end{centering}

\caption{Results for the evolution of the dissipative soliton solution obtained
from numerical simulations for a system with detuning $\Delta_{p}=4\gamma$
and an incoherent pumping rate $P=0.174\gamma$. Insets a) and c)
show the results obtained from direct numerical simulations of the
Maxwell-Bloch(MB) system (equations \ref{eq:waveq} and \ref{eq:Master_equation})
and of the effective Complex Ginzburg-Landau(CGLE) model, respectively,
while b) offers a comparison of both at $z=0.3$ compared with the initial conditions at $z=0$.}
\end{figure}

For a 4-level atom with $\Delta_{p}=4\gamma$ and $P=0.174\gamma$
(which correspond to the values $\delta=0$, $\zeta=1.49$, $\xi=1.55$
and $\epsilon=-5.42$ of the CGLE effective model), equations \ref{eq:ds_parameters}
predict the existence of a dissipative soliton solution of the form
of equation \ref{eq:soliton-1-1} with parameters $A=0.28$, $B=0.087$,
$C=-0.128$ and chirp $d=5.75$. The evolution of this dissipative
soliton was calculated using the MB (see Fig.3a)) and SSFM (see Fig.3c))
solvers. The results obtained are qualitatively similar, despite a
slight difference that can be identified by comparing $\Omega_{p}$
at $z=0.3\mbox{m}$ for both methods (see Fig.3b)). While the results
from the SSFM method preserve the initial intensity profile as it
is expected for a dissipative soliton, the more complete MB method
predicts a small reduction of the intensity amplitude $\left|\Omega_{p}\right|$,
suggesting that this might not be a true dissipative soliton. Therefore,
to clarify if we have indeed a dissipative soliton, we have performed
numerical simulations of non-dissipative solitons given by equation
\ref{eq:soliton-1-1-1} with $a=0.3$ and $a=0.6$(Figure 4). Contrary
to what happens in the dissipative soliton case, the absorption is
now easily identified for small propagation distances, and the pulse
amplitude features an appreciable decay. Moreover, it should be notice
that the case of Figure 4a) corresponds to a soliton solution of the
non-dissipative nonlinear Schrodinger equation with the approximate
amplitude of the dissipative soliton. However, while the amplitude
is approximately the same, the dissipative soliton is usually broader
comparing to the non-dissipative case\cite{DS_sols1996,DS_sols_plasma},
as it can be seen in Fig.4c), which means a different balance between
gain and loss due to spectral filtering processes. Furthermore, in
Figure 4d) is represented the total power that reaches the distance
$z$, normalized to the initial conditions and defined by 
\[
N\left(z\right)\equiv\frac{\int_{-\infty}^{+\infty}\left|\Omega_{p}\left(z,\tau\right)\right|^{2}d\tau}{\int_{-\infty}^{+\infty}\left|\Omega_{p}\left(0,\tau\right)\right|^{2}d\tau}.
\]
It is straightforward to observe that $N$ is conserved in the case
of dissipative soliton conditions, contrary to what happens for the
other initial conditions, where a strong decay is present. Together
with the observations made before, this constitutes a strong evidence
that the type of dissipative soliton solution proposed can be observed
in the atomic system under study. Additional simulations have shown
that these solutions are stable up to distances of $z\approx4\mbox{m}$,
which is consistent with previous results\cite{LAMBDA_DISSIPATIVESOLITONS}
and relevant for a possible experimental observation.

\begin{figure}
\begin{centering}
\includegraphics[width=8.6cm]{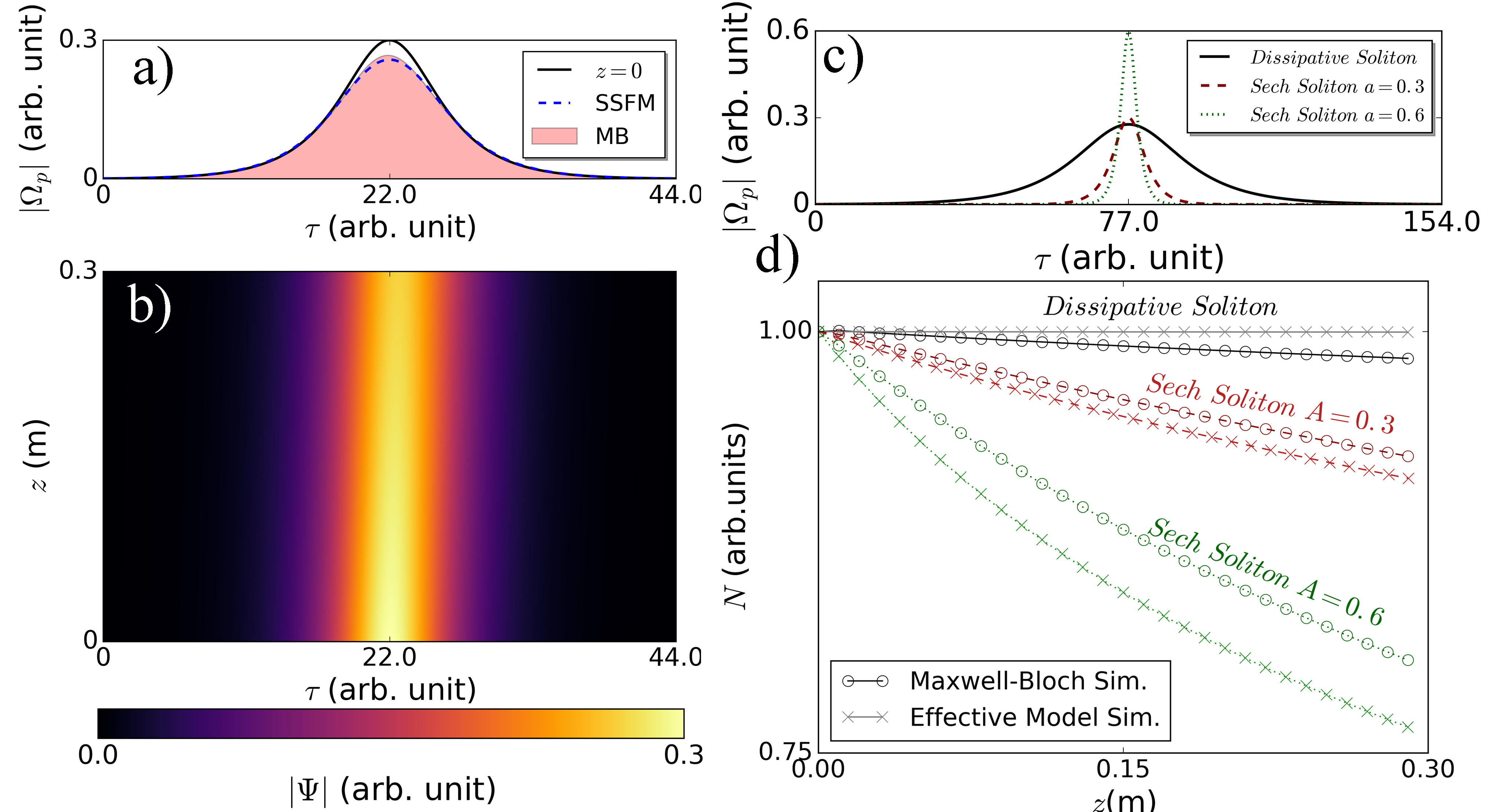}
\par\end{centering}

\caption{Results for the evolution of a soliton solution of form of equation
\ref{eq:soliton-1-1-1} with $a=0.3$, obtained from numerical simulations
for a system with same parameters as Figure 3. Inset a) offers a comparison
of both numerical simulations at $z=0.3\mbox{m}$ compared with the
initial conditions at $z=0$ while b) show the results obtained from
numerical simulations of the Maxwell-Bloch system only. Subfigure
c) show a comparison between the initial conditions for the dissipative
and the non-dissipative type. d) represents the total power that reaches
the distance $z$ for the different types of initial conditions tested.}
\end{figure}

Comparing to the previous work on temporal dissipative solitons in
atomic systems\cite{LAMBDA_DISSIPATIVESOLITONS}, our work introduces
a system to allow a full control of the optical properties and therefore,
to extend the space of parameters of the CGLE to explore in real experiments.
In fact, the assisting incoherent pump allow an easy tuning of the
gain and loss parameters, to an extent previously unavailable. Moreover,
we also stress that our work extends the previous results verifying
the predictions with numerical simulations of not only the effective
model but also of the more complete MB system, which are far more
complete in terms of the dynamics of the optical system in real experiments.
These simulations reveal small but still significant differences,
possibly related to the breakdown of the perturbative approach. Indeed,
the maximum amplitude of the dissipative soliton is $\Omega_{p}\approx0.16\gamma$,
almost of the order of the other parameters of the system, which means
that the dissipative soliton solution is near the boundary of the
limits of the approximation. We have verified that this also happened
in the case explored in previous works\cite{TEMPORALSOLITON_1}, and
therefore constitute an argument that full Maxwell-Bloch simulations
are important to address the full dynamics of the system for future
studies on the existence of dissipative solitons in a real atomic
systems.

\section{Conclusions}

In summary, in this work we have developed a theoretical model for
the propagation of a temporal optical pulse in a $4$-level \emph{N}-type
atomic gas, driven by a continuous-wave electromagnetic field and
an assisting incoherent pump. Employing a multiscale approach, it
is shown that the propagation of a weak optical probe beam inside
the two-dimensional waveguide can be described in terms of an effective
(1+1)-dimensional model in the form of a Complex Ginzburg-Landau Equation,
whose parameters are related with the optical properties of the system.
We have shown that the \emph{N}-type configuration together with the
additional incoherent pumping rate allow a control of the optical
response of the system that extends far beyond the one that was previously
studied, not only in terms of the linear and nonlinear optical properties
but also of gain and loss processes. Testing a family of dissipative
solitons solutions, we have confirmed the possibility of observing
them with computational simulations of both the effective (1+1)-dimensional
model and the full Maxwell-Bloch system of equations, which are far
more complete as it includes most of the dynamical aspects of a real
system. Therefore, extending the space of parameters available and
providing more complete simulations, these results can motivate further
studies on temporal dissipative solitons in atomic systems and can
constitute an useful reference for experimental demonstrations of
the existence of temporal dissipative solitons in these atomic systems.

\begin{acknowledgments}
This work is financed by the European Regional Development Fund (ERDF)
through the Operational Programme for Competitiveness and Internationalisation,
COMPETE 2020 Programme, and by National Funds through the FCT \textendash{} Funda\c{c}\~{a}o para a Ci\^{e}ncia e a Tecnologia
(Portuguese Foundation for Science and Technology), within
Project No. POCI-01-0145-FEDER-032257, as well as by the North Portugal
Regional Operational Programme (NORTE 2020), under the PORTUGAL 2020
Partnership Agreement. N.A.S. is supported by  Funda\c{c}\~{a}o para a Ci\^{e}ncia e a Tecnologia through Grant No. SFRH/BD/105486/2014. This article
is based upon work from COST Action MP1403 \textquotedblleft Nanoscale
Quantum Optics,\textquotedblright{} supported by COST (European Cooperation
in Science and Technology).
\end{acknowledgments}


\bibliographystyle{apsrev4-1}

\end{document}